\PassOptionsToPackage{table}{xcolor}
\documentclass[sigconf,nonacm]{acmart}
\AtBeginDocument{%
  }

\usepackage{xcolor}
\usepackage{colortbl}
\usepackage{tcolorbox}
\usepackage{multirow}
\usepackage{amsmath}
\usepackage{makecell}
\usepackage{booktabs}
\usepackage{tabularx} 
\usepackage{algorithm}
\usepackage{algpseudocode}
\usepackage{listings}

\usepackage{amssymb}
\usepackage{graphicx}

\usepackage[utf8]{inputenc}

\acmConference[]{ }{ }{ }

\begin{document}

\renewcommand\footnotetextcopyrightpermission[1]{}

\title{PUMA: Layer-Pruned Language Model for Efficient Unified Multimodal Retrieval with Modality-Adaptive Learning}

\author{Yibo Lyu \quad Rui Shao$^{\dagger}$ \quad Gongwei Chen \quad Yijie Zhu \quad Weili Guan \quad Liqiang Nie$^{\dagger}$ \\
\texttt{Harbin Institute of Technology, Shenzhen} \\
\texttt{weberlv1b@gmail.com        \{shaorui, nieliqiang\}@hit.edu.cn} \\
\url{https://github.com/JiuTian-VL/PUMA}
}

% \affiliation{%
%   \institution{
%   Harbin Institute of Technology, Shenzhen}
%   % \authornote{Corresponding author}
% }
% \email{weberlv1b@gmail.com {shaorui, nieliqiang}@hit.edu.cn}

\thanks{$^{\dagger}$Corresponding author.}
\renewcommand{\shortauthors}{ }

\begin{abstract}
  As multimedia content expands, the demand for unified multimodal retrieval (UMR) in real-world applications increases. Recent work leverages multimodal large language models to tackle this task.
  However, the large number of parameters leads to high training resource demands and low inference efficiency. To address this issue, we propose the PUMA: Layer-\textbf{P}runed Language Model for Efficient \textbf{U}nified \textbf{M}ultimodal Retrieval with Modality-\textbf{A}daptive Learning, an efficient approach to enhancing the unified retrieval capabilities from both structure and learning perspectives:
  \textbf{1)} From the perspective of model structure, to retain the most retrieval-relevant components within MLLMs, we analyze and propose \textbf{Layer-Pruned Self-Distillation} approach. It structurally prunes the model by preserving only the shallow layers, substantially reducing the parameters of MLLM. Moreover, we use self-distillation to mitigate the representational degradation caused by pruning. It reuses the feature from dropped deep layers as the teacher signal, where the supervised signal enables the retrieval embedding token to efficiently inherit effective representational capacity, resulting in a more compact model.
  \textbf{2)} From the perspective of model learning, to mitigate representation degradation caused by rapid convergence during multimodal contrastive learning, we propose \textbf{Modality-Adaptive Contrastive Learning} Loss (MAC-Loss). It adaptively separates in-batch negative candidate samples into harder intra-modality and simpler inter-modality groups based on each query’s target modality. Assigning each group a temperature coefficient with different strategies enables each query to adaptively focus on challenging in-batch negatives, reducing the resource demands of multimodal contrastive learning.
  Experiments demonstrate that our approach achieves double efficiency, significantly reduces resource consumption while maintaining most of the performance.

\end{abstract}

\maketitle

\section{Introduction}
\begin{figure}[htbp]
  \centering
  \vspace{5pt}
  \includegraphics[width=1.0\linewidth]{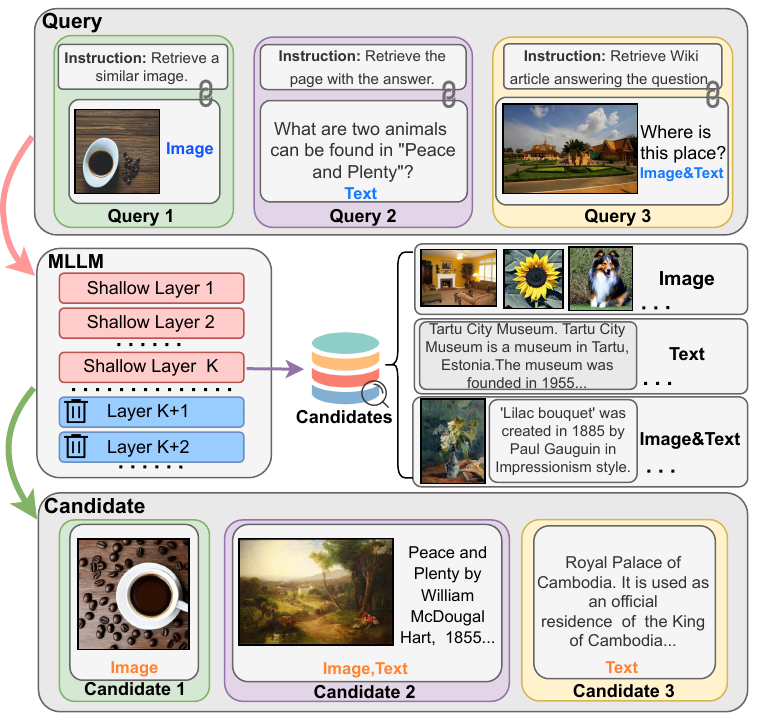}
  \caption{The pipeline of PUMA. We propose an efficient approach that enables MLLMs to perform the UMR task, allowing the model to accept arbitrary-modality input and retrieve from mixed-modality candidates following instructions. Our pruning strategy significantly improves efficiency with comparable performance to the original MLLM.}
  \Description{}
  % \vspace{-5pt}
  \label{intro}
\end{figure}

Multimodal retrieval, a core task in information retrieval (IR), aims to retrieve relevant content across different data modalities \cite{lee2018stacked, zhang2018deep, li2021align}. A more general and challenging setting is Unified Multimodal Retrieval (UMR) \cite{wei2024uniir, lin2025mmembed}, where both queries and candidates can involve arbitrary modality combinations. While CLIP-based models perform well in fixed-modality input scenarios \cite{radford2021learning, li2022blip, luo2022clip4clip}, they struggle to integrate more complex multimodal scenarios. In contrast, Multimodal Large Language Models (MLLMs) \cite{liu2023visual, li2024llava, wang2024qwen2, chen2024lion, zhang2025falcon, shen2024mome}, pretrained on large-scale image-text data, excel in multimodal understanding and real-world knowledge \cite{li2024optimus, li2025optimus, li2025lion, shao2024detecting, shao2023detecting}, making them ideal for UMR. Despite their autoregressive training, studies on the MTEB benchmark \cite{li2023towards, wang2024improving, chen-etal-2024-m3, muennighoff2023mteb} show that strong language understanding could enhance retrieval capability. Recent works \cite{liu2025lamra, lin2025mmembed, huang2025joint} further demonstrate that applying MLLMs to UMR outperforms the CLIP-based methods.

However, using large models (e.g., 7B or more) for retrieval remains inefficient in both training and inference, often requiring substantial computational resources and leading to increased costs for downstream tasks, which poses challenges for real-world deployment. To address this issue, as illustrated in Figure~\ref{intro}, we propose \textbf{PUMA}: a Layer-Pruned Language Model for Efficient Unified Multimodal Retrieval with Modality-Adaptive Learning framework, which achieves a more efficient UMR from the \textbf{structure and learning perspectives}. \textbf{1) From the model structure perspective.} To improve model efficiency, we aim to reduce the number of parameters by identifying and retaining only the components most relevant to retrieval. Recent studies on interpretability and layer functionality in MLLM \cite{zhang2024redundancy, kaduri2024s, gromov2403unreasonable, sawtell2024lightweight, fischer2024large, liu2024pruning} have shown a similar pattern across MLLMs: in Visual Question Answering (VQA) task, fine-grained multimodal integration primarily occurs in the shallow layers, while deep layers are mainly responsible for next-token prediction. These works imply that shallow layers are more valuable for the retrieval task, but their analysis is limited to VQA task. We extend this investigation to retrieval tasks and observe a similar phenomenon in unified retrieval as in VQA task, enabling efficiency through model structure lightweighting via layer pruning.
 
However, layer-pruning still damages the representation capability of MLLMs \cite{gromov2403unreasonable, yang2024laco, zhang2024finercut}, as the primary role of shallow layers is to aggregate information for subsequent layers rather than directly performing semantic representation. Previous work typically discards layers directly, necessitating more training to recover the resulting degradation \cite{gromov2403unreasonable}. To address this issue, we incorporate \textbf{layer-pruned self-distillation}, where the original model (before pruning) and the pruned model are treated as the teacher and student model, respectively. Specifically, we use the embedding feature from the original model to supervise the retrieval token from the pruned layers. This allows the shallow representations of the pruned model to benefit from the rich representation features of the original model, effectively inheriting representational capacity while significantly reducing the number of parameters. Meanwhile, it is jointly pretrained with contrastive loss, enabling the pruned model to quickly adapt to the UMR task.

\textbf{2) From the model learning perspective.} We find that the inherent gap between different modality embeddings in UMR often leads to easy negative samples, causing rapid convergence and degraded representation under InfoNCE loss \cite{oord2018representation, chen2020simple}. Increasing batch size \cite{he2020momentum, chen2021empirical} and hard negative sampling \cite{robinson2021contrastive, kalantidis2020hard} can raise negative sample difficulty, but both significantly increase computational cost, especially for MLLM-based models.
To address this issue, we propose a \textbf{modality-adaptive contrastive learning} loss (MAC-Loss) that performs hard negative sampling without introducing extra cost.  
Through dimensionality reduction and visualization, we observe that separating in-batch samples by modality naturally highlights the harder negatives within each batch.
Motivated by this, MAC-Loss adaptively splits in-batch negatives into harder intra-modality and simpler inter-modality groups based on each query's target modality. By explicitly separating intra- and inter-modality negatives, the model is better positioned to identify and prioritize the harder negatives by modality during training. To implement this focus, we assign different temperature strategies to intra- and inter-modality negatives during training. This guides the model to pay greater attention to the challenging intra-modality negatives within each batch. Our contributions can be summarized as follows:
\begin{itemize}
\item We analyze and propose layer-pruned self-distillation that leverages the inherent capabilities of the original MLLM, while layer-pruning can get a more compact and efficient model for UMR from the model structural perspective.
\item We design a modality-adaptive contrastive learning loss to achieve in-batch hard negative sampling, further reducing the dependency on computational resources from the model learning perspective. 
\item Experiments demonstrate that both method is highly effective, significantly reducing training, inference costs while preserving most of the performance.
\end{itemize}

\section{Related Work}
\subsection{Multimodal Representation Learning}
Many methods have been explored for multimodal representation learning. Models like CLIP \cite{radford2021learning} and ALIGN \cite{jia2021scaling} have achieved impressive results through large-scale image-text contrastive learning. BLIP \cite{li2022blip} further enhances cross-modal representation capabilities by integrating contrastive learning, generative pretraining, and image-text matching. Many other representation methods have also been developed \cite{zhai2023sigmoid, luo2022clip4clip, shao2019multi}. For the retrieval domain, UniIR \cite{wei2024uniir} integrates datasets from multiple retrieval scenarios, and training CLIP across diverse modalities enhances its generalization capability. 
However, CLIP-based models still face limitations in handling more complex tasks or flexible input formats (e.g., videos or interleaved image-text). MLLMs offer a promising alternative. E5V \cite{jiang2024e5} shows that well-designed prompts can guide MLLMs to align images and text within the hidden space and perform retrieval tasks. Based on this, recent works like MMEmbed \cite{lin2025mmembed} and LamRA \cite{liu2025lamra} leverage the strong multimodal understanding capabilities of MLLMs to generate unified retrieval embeddings extending UniIR \cite{wei2024uniir}. Despite their advantages, MLLM-based models often suffer from high computational costs. In this paper, we try to address the efficiency challenges of MLLM-based retrieval models.

\subsection{Efficient Multimodal Language Model}
Following the success of large language models \cite{team2023gemini, grattafiori2024llama, jiang2024mixtral}, multimodal large models have also attracted extensive attention and development \cite{li2024llava, xie2025gui, chen2025SimpAgent, listar, li2025generative, qin2024uno}. However, their efficiency is a major concern due to the large number of parameters.
To address this, recent research has explored ways to improve the efficiency of MLLMs. Some approaches observe that visual tokens become redundant after a few layers and they improve efficiency at the token-level by dynamically dropping them \cite{chen2024image, shang2024llava, zhang2024token}, or by compressing them using learnable tokens \cite{ye2024voco, wen2024efficient}. While others focus on the layer-level by pruning parts of the model to reduce parameters \cite{gromov2403unreasonable, sawtell2024lightweight, fan2024not}. 
For retrieval tasks, methods that only compress visual tokens are not suitable for handling text-only embeddings. In this work, we aim to improve the efficiency of UMR at the layer-level. We adopt a layer-pruning strategy to reduce model size directly, improving both training and inference efficiency across all modalities.

\subsection{Representation Learning}
Knowledge distillation has proven effective in representation learning, where a teacher model guides a student model during training. FitNets \cite{romero2014fitnets} introduced using intermediate features from the teacher to supervise the student, while CLIPPING \cite{pei2023clipping} proposed a hierarchical alignment approach that aligns the student’s intermediate layers with the teacher’s. Since layer pruning will disrupt the representational continuity of large models, distilling the representation feature from the original model helps recover performance to shallow layers and enables training to be more efficient.

\begin{figure*}[htbp]
  \centering
  \includegraphics[width=0.99\linewidth]{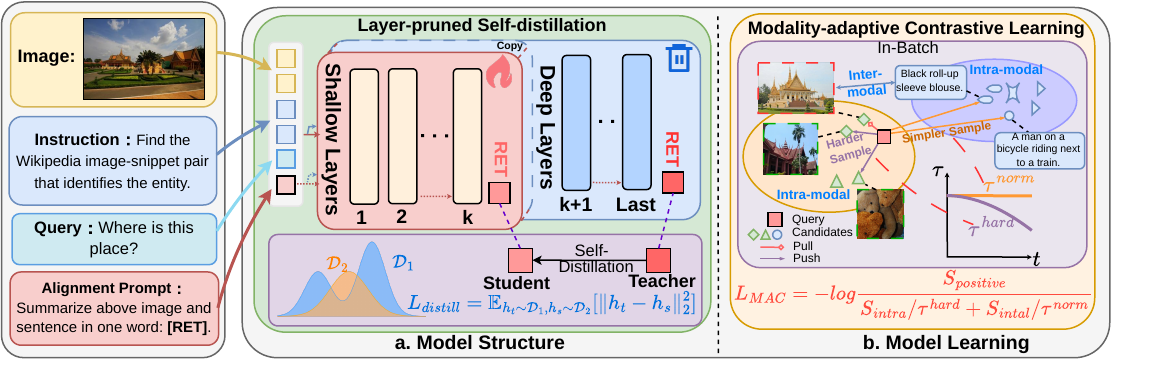}
  \caption{Overview of the PUMA framework. Our method comprises two key components from both the model architecture and learning perspective: layer-pruned self-distillation and modality-adaptive learning. (a). We propose a layer-pruned self-distillation approach that reduces model parameters while preserving performance through distillation for [RET] token. (b). Modality-adaptive learning loss divides in-batch samples for each query into inter- and intra-modality groups, applying different temperature strategies to enable adaptive hard negative sampling based on the query’s target modality.}
  \label{arch}
  \vspace{-8pt}
  \label{overview}
\end{figure*}

Contrastive learning is another core technique in representation and self-supervised learning, encouraging the model to pull positive pairs closer and push negative ones apart. MoCo \cite{he2020momentum} uses a momentum encoder and dynamic dictionary to support scalable contrastive learning. SimCLR \cite{chen2020simple} boosts performance through a simple framework and strong data augmentations. FlatNCE \cite{chen2021simpler} demonstrates that overly simple negatives cause the InfoNCE \cite{oord2018representation} loss to vanish quickly. To address this, we introduce a modality-adaptive learning loss that incorporates hard negative sampling adaptively for each query in-batch without extra sampling computation and reduces the reliance on large batch and negative sampling to some extent.

\section{Preliminary} 
To extract embedding representation in the MLLM, we follow the approach of previous works \cite{jiang2024e5, liu2025lamra, lin2025mmembed, huang2025joint}, using a well-designed prompt to constrain a single word, this word position will effectively aggregate multimodal information. Specifically, our prompt is:

\begin{tcolorbox}
\texttt{Question: \{$I^{image}$\} <Instruction> \{$I^{query}$\}. \textbackslash nSummary above \{image / sentence / image and sentence\} in one word: [RET].}
\end{tcolorbox}
Here, $I^{image}$ represents inputs with image, $I^{query}$ represents inputs with query text. The input modalities are flexibly combined, and the following instructions are adapted accordingly. [RET] denotes a special token registered in the LLM. We use the hidden state at this special token position as the retrieval embedding. In this paper, our experiments are mainly conducted on Qwen2-VL \cite{wang2024qwen2}, which has recently demonstrated strong performance in multimodal alignment.

We use the same training strategy and dataset as LamRA \cite{liu2025lamra}. In the pre-training stage, the model is fine-tuned on a text-to-text retrieval dataset. In the second stage, we perform instruction tuning on the M-BEIR training set \cite{wei2024uniir}, which includes diverse retrieval tasks to enhance the model's unified retrieval capability. Specially, for the layer of MLLM $L= \{{L_1, L_2, ..., L_{last}} \}$, we only extract the hidden state up to a certain layer ${L_k}$ and take the hidden state at the [RET] position, denoted as $h_k$, as the retrieval embedding. 

Given a query $q$ in any modality, including images, text, or interleave image-text pairs, etc. Our objective is to retrieve the most relevant response from a candidate pool $C = \{c_{1}^i, c_2^i, ..., c_i^t, c_{i+1}^t, ..., c_n^{i,t}\}$, where $c_j^m$ denotes the $j$-th candidate in modality $m$. The candidates span arbitrary modalities. We first obtain the embeddings of $q$ and the candidates in $C$ using the designed prompt. Subsequently, we compute their cosine similarity and select the top k candidates from $C$ with the highest semantic relevance.

\section{PUMA}
In this section, we provide more details about PUMA. As shown in Figure \ref{overview}, our method combines two key components: layer-pruned self-distillation and modality-adaptive contrastive learning, focusing on model structure and learning perspectives. We provide an analysis for the retrieval domain in Section \ref{4.1}, followed by an explanation of the two methods in Sections \ref{4.2} and \ref{4.3}.

\subsection{Analysis of Shallow Layer in UMR Task}
\label{4.1}
A growing body of research has demonstrated that leveraging the shallow layers of LLMs is highly effective for some downstream tasks such as security monitoring, sentiment analysis, and even text retrieval \cite{gromov2403unreasonable, sawtell2024lightweight, fischer2024large}. These findings suggest that utilizing shallow layers for certain downstream applications is efficient. Provided us with some successful precedents.

On the other hand, to accomplish the UMR task, we expect the model to possess the following capabilities: (1) effective interaction and fusion of multimodal information; and (2) the ability to embed multimodal information in some token. Previous studies on different MLLMs have provided encouraging insights for shallow layers:

(i) Several recent works \cite{chen2024image, zhang2025llava, zhang2024redundancy} have explored attention mechanisms in MLLMs, revealing that attention between image and text modalities is dense in the shallow layers but becomes increasingly sparse in deeper layers. \textbf{Support that shallow layer may mainly involve information interaction in MLLM.}

(ii) Some token compression methods \cite{chen2024image, wen2024efficient, ye2024voco} have shown that discarding image tokens after a few shallow layers or applying some learnable tokens to gather image information has a negligible impact on the final results. \textbf{Support that shallow layers could fuse information in some token.}

Motivated by promising findings in VQA, we investigate whether shallow layers still exhibit similar beneficial properties after training on the unified retrieval dataset \cite{wei2024uniir}. This allows us to assess whether the methods validated in VQA can be transferred to UMR.

\begin{figure}[htbp]
  \vspace{-8pt}
  \centering
  \includegraphics[width=1.0\linewidth]{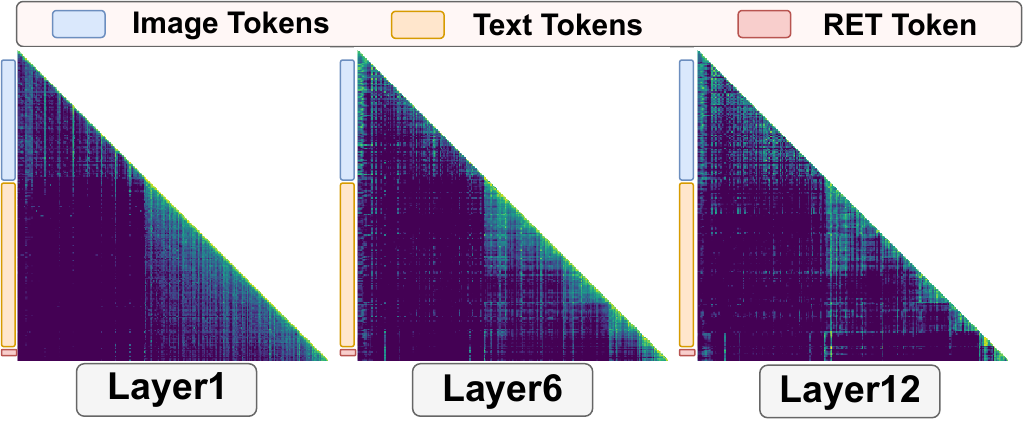}
  \caption{Visualization of the attention map of retrieval embeddings in MLLMs. The legend indicates the modality of each token, aligned from left to right with the bottom row.}
  \label{fig3}
  \vspace{-8pt}
\end{figure}

As shown in Figure \ref{fig3}, following the previous studies \cite{chen2024image, zhang2025llava}, we visualize the shallow layer attention mechanisms of Qwen2-VL after contrastive learning. 
Similar to observations in the VQA task, the attention maps show that self-modality interactions within both image and text remain dense in the shallow layers. As the model progresses through deeper layers, the retrieval ([RET]) token appears to attend more to specific image and text tokens. These findings resemble those observed in VQA tasks and also show patterns that may benefit retrieval. This suggests that the function of shallow layers may not differ significantly between retrieval and VQA, and therefore, we try to apply the pruning strategies that have proven effective in VQA to the UMR task.

\subsection{Layer-Pruned Self-Distillation}
\label{4.2}
Although pruning layers can make the model more lightweight and efficient, it still results in a performance decrease. The primary role of shallow layers in MLLMs is to aggregate information for the next layer. Directly extracting the obtained feature is essentially equivalent to omitting the decoding process from ${L_k}$ to ${L_{last}}$. As the hidden state moves further from the final layer, its ability to express meaningful information decreases. As a result, pruning while retaining only the shallow layers causes performance degradation. To qualitatively characterize this phenomenon, we use ${\phi(\cdot)}$ to denote the capability of effective information representation, which can be formulated as: 
\begin{equation}
    b^-(\phi(h_k), len(L_{last}-L_k)),
\end{equation}
where ${b^-}(\cdot, \cdot)$ indicates a negative correlation, ${h_k}$ indicates [RET] position hidden state of layer ${L_k}$.

The retrieval embeddings from truncated shallow layers need to restore the representational capacity similar to that of the original model while also learning effective retrieval-specific features to accomplish the UMR task. Fully enhancing the retrieval capability of shallow layers may require more training data or additional training stages. To avoid this overhead, we propose a layer-pruned self-distillation approach that reuses the representational power of the pruned layers. The final decoded features act as auxiliary supervised signals to guide shallow layers in learning effective information representations more efficiently.

Therefore, we use feature-level knowledge distillation (KD) to implement this. Specifically, the retrieval hidden-state from the last layer, $h_t$, serves as the teacher, while this from the shallow layer, $h_s$, acts as the student. We use Mean Squared Error (MSE) loss, a common loss for feature distillation, to help align their representation features.
Given a query-candidate pair ${{q}, {c}}$, their embeddings are denoted as ${{h^q}, {h^c}}$. The last layer hidden states are obtained via $MLLM_{1 \to last}(q, c) \to {h_t^q, h_t^c}$, while the shallow layer states are $MLLM_{1 \to k}(q, c) \to {h_s^q, h_s^c}$. The self-distillation loss is formulated as:
\begin{equation}
    L_{self-distill} = \mathbb{E}_{ h_t \sim \mathcal{D_\text{1}}, h_s \sim \mathcal{D_\text{2}}} \left[ \| h_t^q - h_s^q \|_2^2 + \| h_t^c - h_s^c \|_2^2 \right],
\end{equation}
where ${\mathcal{D_\text{1}}}$, ${\mathcal{D_\text{2}}}$ represent the feature distributions of the teacher and student, respectively. We assist the shallow layers in learning the original model’s feature representations by computing the loss for both the queries and candidates.

On the other hand, to learn retrieval representations, we also employ the InfoNCE loss \cite{oord2018representation} for contrastive learning, which is defined as:
\begin{equation}
    \mathcal{L}_{contrastive} = -\frac{1}{N} \sum_{i=1}^{N} \log \frac{\exp(\text{sim}(h^{i,q}, h^{i,c}) / \tau)}
{\sum_{n=1}^{N} \exp(\text{sim}(h^{i,q}, h^{n,c}) / \tau)},
\end{equation}
where $sim(\cdot, \cdot)$ indicates cosine similarity, $\tau$ indicates temperature coefficient. The remaining candidates are considered as negative samples within the same batch.
The loss for the first stage can be expressed as:
\begin{equation}
    \mathcal{L}_{pretraining} = \alpha_1 \mathcal{L}_{contrastive} + \alpha_2 \mathcal{L}_{self-distill}.
\label{distill equation}
\end{equation}
For ${L_{pretraining}}$, the ${L_{contrastive}}$ component plays the primary role, as the goal of pre-training is to improve the model’s retrieval capability, which relies more on contrastive learning. The ${L_{self-distill}}$ term is used as an auxiliary to help efficient training during the pre-training stage only. Setting similar values for ${\alpha_1}$ and ${\alpha_2}$ weakens the model’s ability to distinguish between samples, thereby harming retrieval performance.

\subsection{Modality-Adaptive Learning}
\label{4.3}
\begin{figure}[htbp]
\vspace{-8pt}
  \centering
  \includegraphics[width=1.0\linewidth]{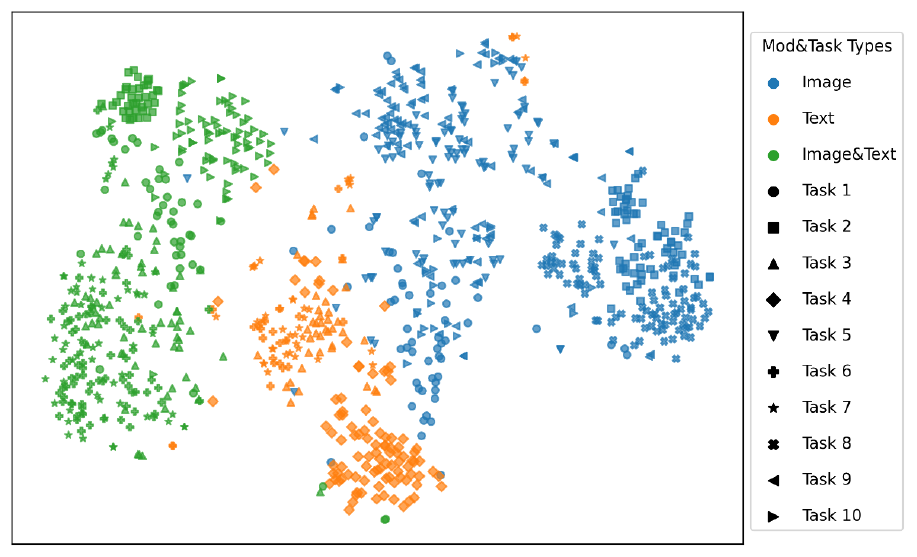}
  \caption{Visualization of data distribution. We use t-SNE for dimensionality reduction to visualize the random samples from the M-BEIR, where different colors indicate candidate modalities and different shapes represent different tasks.}
  \label{tSNE}
  \vspace{-15pt}
\end{figure}

In the second stage, we continue training on the M-BEIR dataset \cite{wei2024uniir}, which includes 8 tasks across 10 datasets. We highlight that directly conducting contrastive learning training on such dataset presents several challenges. To simplify, we further refine the InfoNCE loss as follows:
\begin{equation}
\begin{split}
\mathcal{L}_{i}
&= -\log \frac{S_i}{\sum_n S_n} =  -\log (1+ \sum_{n \neq i}\frac{S_n}{S_i}),
\end{split}
\label{contrastive}
\end{equation}
where $L_i$ means contrastive loss for each sample in-batch, $S_i$ indicates the cosine similarity score between query and positive candidate ${c^i}$, $S_n$ indicates the cosine similarity score with other in-batch candidate samples ${c^n}$. Through Equation \ref{contrastive}, we observe that selecting appropriate negatives $S_n$ is crucial for in-batch contrastive learning. If the negative samples are too simple, the similarity score between the query and in-batch candidate samples may become too low, formulated as ${S_n\to 0}$, causing the contrastive loss to quickly approach zero \cite{chen2021simpler}. This will hinder the model from learning meaningful representations, leading to poor representation quality or even "representation collapse". Unfortunately, Figure \ref{tSNE} reveals inherent disparities in embeddings across different candidate modalities. During mixed-modality training, this accelerates the learning process by introducing more easily distinguishable negatives from different modalities within each batch. For example, image candidates x, y, z are easier to distinguish from a positive text candidate $\gamma$, leading to potentially premature convergence.

To address this issue, two solutions can be considered: (1) increasing the batch size, which naturally introduces more challenging negative samples, and (2) performing hard negative sampling to deliberately select difficult negative samples. While increasing the batch size can improve training stability to some extent, it comes with a significant GPU cost, particularly when using an MLLM as the backbone. On the other hand, in the second training stage, the M-BEIR dataset includes a candidate pool of 5.6 million, and applying hard negative sampling in the global candidate pool would also result in substantial sampling overhead. As a result, both approaches lead to significant computational costs. The dataset sampling strategy will lead to the long-tail problem and unstable training gradients in M-BEIR, both the data quantity distribution and feature distribution are imbalanced and unstable. These all highlight the need for a more efficient strategy specifically designed for MLLM-based models, especially mixed-modality training.

To address this challenge, we introduce modality-adaptive contrastive learning loss tailored for mixed-modality training during instruction tuning. Based on the observation from Figere \ref{tSNE}, we can further decompose ${S_n}$ by modality adaptively for each query ${q_i}$, formulated as: 
\begin{equation}
    S_n = 
        \begin{cases}
        S_{intra}, & \text{if } c_i^{m} = c_j^{m} \\
        S_{inter}, & \text{if } c_i^{m} \neq c_j^{m}
        \end{cases},
\end{equation}
where ${c_i^m}$ means the modality of the candidate ${i}$. We can adaptively partition all in-batch negative samples for each query based on its target candidate modality, enabling hard negative sampling without incurring additional computational overhead. and the denominator of the InfoNCE loss can be decomposed as:
\begin{equation}
\mathcal{L_{MAC}} = -\log \frac{\exp(S_p / \tau)}
{\sum_{m=1}^{M} \exp(S_m^{intra} / \tau^{hard}) + 
\sum_{n=1}^{N} \exp(S_n^{inter} / \tau^{norm})}.
\label{mod loss}
\end{equation}

To encourage the model to focus more on harder in-batch candidates in group ${S^{intra}}$, we adjust the temperature coefficient ${\tau}$ after partitioning the contrastive loss, guiding the model to pay greater attention to the ${S^{intra}}$. Specifically, we assign a separate temperature coefficient $\tau$ to the intra-modality group $S^{intra}$ and gradually decay it during training. This dynamic adjustment increases the sharpness of the similarity distribution within each batch, effectively amplifying the contribution of harder negative samples and encouraging the model to better discriminate between subtle differences during mixed-modality optimization. Meanwhile, the gradual adjustment helps balance the focus between intra- and inter-modality samples, preventing the model from overemphasizing hard intra-modality examples.
\begin{equation}
    \tau^{hard} = \tau_{0} \cdot e^{-\lambda t}, \tau^{norm} = \tau_0
\end{equation}
where $\lambda$ represents the decay sparsity and $t$ denotes the current iteration number. During training, we utilize our MAC-loss to replace the conventional contrastive learning approach.

\vspace{-3pt}
\begin{algorithm}[h]
    \caption{\textbf{Pseudocode of MAC-Loss in Pytorch-like Style.}}
    \label{tab:oemb_pseudocode}
    \begin{lstlisting}[language=Python, basicstyle=\footnotesize\ttfamily, numbers=none, aboveskip=0pt, belowskip=0pt]
# query_inputs: query inputs for MLLM
# cand_inputs: candidate inputs for MLLM
# modality_target: target candidate modality of each query
# norm_temp: contrastive learning temperature 
# lambda: decay sparsity
q_embed, c_embed = model(**query_inputs), model(**cand_inputs)
q_gather, c_gather = dist_gather(q_embed), dist_gather(c_embed) # gather data from other GPUs
q_ret, c_ret = F.normalize(q_gather, p=2, dim=-1), F.normalize(c_gather, p=2, dim=-1)
scores = torch.matmul(q_ret, c_ret.transpose(0, 1))
hard_temp = round(norm_temp * math.exp(-lambda * (current_epoch / total_epochs), 3)
modality_matrix = (modality_target.unsqueeze(0) == modality_target.unsqueeze(1))
matrix_temp = torch.where(modality_matrix, hard_temp, norm_temp)
scores = scores / matrix_temp
    \end{lstlisting}
\label{algorithm}
\end{algorithm}
\vspace{-5pt}
Algorithm \ref{algorithm} presents the pseudocode for our MAC-Loss in a pytorch-like manner. Our learning loss function will not introduce additional hard negative sampling; instead, it operates effectively within a standard in-batch contrastive loss setting.

% \multicolumn{8}{c}{Single Modal}
\begin{table*}[ht!]
\setlength{\tabcolsep}{1pt}
\small
\centering
\caption{Retrieval Recall on the M-BEIR benchmark \cite{wei2024uniir}.
We group the eight tasks into three types based on input and output modalities. "Single": both input and output are unimodal. "Mixed": either input or output is multimodal. "Multi": both input and output are multimodal. ${q_t}$ and ${c_t}$ denote text queries and candidates; ${q_i}$ and ${c_i}$ denote image queries and candidates. We compare our model with LamRA-RET, where both parameters are smaller than 4B.}
\vspace{-0.1in}
\setlength{\tabcolsep}{1.5mm}{
% \resizebox{\textwidth}{!}{%
\begin{tabular}{lc@{\hspace{0.15cm}}c@{\hspace{0.1cm}}c@{\hspace{0.1cm}}c@{\hspace{0.1cm}}c@{\hspace{0.15cm}}c@{\hspace{0.1cm}}c@{\hspace{0.1cm}}c@{\hspace{0.1cm}}c@{\hspace{0.1cm}}c@{\hspace{0.1cm}}c@{\hspace{0.1cm}}c@{\hspace{0.1cm}}c@{\hspace{0.15cm}}c@{\hspace{0.1cm}}c@{\hspace{0.1cm}}c@{\hspace{0.1cm}}c@{\hspace{0.1cm}}}
\toprule
 \multirow{3}{*}{\textbf{Models}} &  \multicolumn{8}{c}{\textbf{Single Modal}} & \multicolumn{6}{c}{\textbf{Mixed Modal}} & \multicolumn{2}{c}{\textbf{Multi Modal}} & \multirow{3}{*}{\textbf{Avg}} \\
 \cmidrule(lr){2-9} \cmidrule(lr){10-15} \cmidrule(lr){16-17}
 & \multicolumn{3}{c}{$q_t \to c_i$} & $q_t \to c_t$ & \multicolumn{3}{c}{$q_i \to c_t$} & $q_i \to c_i$ & \multicolumn{2}{c}{$q_t \to (c_i,c_t)$} & \multicolumn{2}{c}{$(q_i, q_t) \to c_t$} & \multicolumn{2}{c}{$(q_i, q_t) \to c_i$} & \multicolumn{2}{c}{$(q_i, q_t) \to (c_i, c_t)$} \\
 \cmidrule(lr){2-4} \cmidrule(lr){5-5} \cmidrule(lr){6-8} \cmidrule(lr){9-9} 
 \cmidrule(lr){10-11} \cmidrule(lr){12-13} \cmidrule(lr){14-15} \cmidrule(lr){16-17}
 & VN & COCO & F200K & WebQA & VN & COCO & F200K & Nights & EDIS & WebQA & Oven & InfoS & FIQ & CIRR & Oven & infoS & \\
\midrule
\multicolumn{18}{c}{\textit{Zero-shot}} \\
\midrule
CLIP \cite{radford2021learning} & 43.3 & 61.1 & 6.6 & 36.2 & 41.3 & 79.0 & 7.7 & 26.1 & 43.3 & 45.1 & 24.2 & 20.5 & 7.0 & 13.2 & 38.8 & 26.4 & 32.5 \\
SigLip \cite{zhai2023sigmoid} & 30.1 & 75.7 & 36.5 & 39.8 & 30.8 & 88.2 & 34.2 & 28.9 & 27.0 & 43.5 & 29.7 & 25.1 & 14.4 & 22.7 & 41.7 & 27.4 & 37.2 \\
BLIP \cite{li2022blip} & 16.4 & 74.4 & 15.9 & 44.9 & 17.2 & 83.2 & 19.9 & 27.4 & 26.8 & 20.3 & 16.1 & 10.2 & 2.3 & 10.6 & 27.4 & 16.6 & 26.8 \\
BLIP2 \cite{li2023blip} & 16.7 & 63.8 & 14.0 & 38.6 & 15.0 & 80.0 & 14.2 & 25.4 & 26.9 & 24.5 & 12.2 & 5.5 & 4.4 & 11.8 & 27.3 & 15.8 & 24.8 \\
Qwen2VL \cite{wang2024qwen2} & 9.3 & 55.1 & 5.0 & 42.0 & 5.4 & 46.6 & 4.0 & 21.3 & 26.2 & 9.4 & 21.4 & 22.5 & 4.3 & 16.3 & 43.6 & 36.2 & 23.0 \\ 
\midrule
\multicolumn{18}{c}{\textit{Supervised Clip-Based Model}} \\
\midrule
$BLIP_{SF}$ \cite{wei2024uniir} & 23.4 & 79.7 & 26.1 & 80.0 & 22.8 & 89.9 & 28.9 & 33.0 & 50.9 & 79.8 & 41.0 & 22.4 & 29.2 & 52.2 & 55.8 & 33.0 & 46.8 \\
$CLIP_{SF}$ \cite{wei2024uniir} & 42.6 & 81.1 & 18.0 & 84.7 & 43.1 & 92.3 & 18.3 & 33.0 & 50.9 & 78.7 & 45.5 & 27.9 & 24.4 & 44.6 & 67.6 & 48.9 & 50.6 \\
\midrule
\multicolumn{18}{c}{\textit{Supervised MLLM-Based Model(<4B)}} \\
\midrule
LamRA-Ret \cite{liu2025lamra} & 30.8 & 78.8 & 25.1 & 82.5 & 31.2 & 88.9 & 27.1 & 28.7 & 54.3 & 77.8 & 51.1 & 44.2 & 28.9 & 47.7 & 72.3 & 60.8 & 51.8 \\
\rowcolor{blue!10}
PUMA & 34.4 & 79.9 & 25.4 & 86.1 & 34.2 & 90.5 & 27.9 & 32.0 & 57.2 & 78.2 & 53.0 & 47.3 & 30.4 & 49.1 & 74.1 & 65.0 & \textbf{54.1} \\

\bottomrule
\label{main table}
\end{tabular}
}
\vspace{-7pt}
\end{table*}

\section{Experiment}

\begin{table*}
  \vspace{-5pt}
  \caption{Comparison of efficiency with larger MLLM-based retrievers. We present detailed results on UMR models' retrieval capability and efficiency. Inference speed refers to the number of samples processed per second during inference, while FLOPs (floating-point operations) measure the computational cost and are commonly used to evaluate the efficiency of LLMs. MMEmbed* with a different backbone and training setup, so we omit detailed comparison here.}
  \setlength{\tabcolsep}{2pt}
  \begin{tabular}{c c | c c c| c c c c }
    \toprule
    Models & Backbone & Single & Mixed & Multi & LLM Parameter & Training Sources & FLOPs $\downarrow$ & Inference Speed $\uparrow$ \\
    \midrule
    MMEmbed* \cite{lin2025mmembed} & LLaVA-Next & 50.9 & 52.3 & 60.9 & 7B & 8*80G & - & -\\
    \midrule
    LamRA-Ret \cite{liu2025lamra} & Qwen2-VL & 53.6 & 55.2 & 69.8 & 7B & 16*80G & 7.36 & 59.0\\
    \rowcolor{blue!10}
    PUMA & Qwen2-VL (Pruned) & 51.3 \textcolor[rgb]{0.8, 0.3, 0.3}{\small$\downarrow$ 2.3} & 52.6 \textcolor[rgb]{0.8, 0.3, 0.3}{\small$\downarrow$ 2.6} & 69.6 \textcolor[rgb]{0.8, 0.3, 0.3}{\small$\downarrow$ 0.2} & \textbf{3B}\textcolor[rgb]{0.0, 0.5, 0.0}{\small$\downarrow$ \textbf{52.6\%}}  & \textbf{4*80G} \textcolor[rgb]{0.0, 0.5, 0.0}{\small$\downarrow$ \textbf{4x}} & \textbf{3.48} \textcolor[rgb]{0.0, 0.5, 0.0}{\small$\downarrow$ \textbf{52.7\%}}  & \textbf{115.5} \textcolor[rgb]{0.0, 0.5, 0.0}{\small$\uparrow$ \textbf{95.8\%}} \\
    % LamRA-Ret & Qwen2VL & 49.6 & 50.6 & 66.5 & 2B & 0.82 & 59.0 \\
  \bottomrule
\label{efficient}
\end{tabular}
\vspace{-10pt}
\end{table*}

\subsection{Training Setup}
\textbf{Datasets.}~The training process includes two stages using LoRA \cite{hu2022lora}. In the first stage, we train the model on the text-to-text retrieval task on the Natural Language Inference (NLI) dataset \cite{gao2021simcse}. In the second stage, we continue instruction tuning on the M-BEIR dataset \cite{wei2024uniir}, enabling the model to develop UMR capability.
For evaluation, we use Recall@k to measure retrieval performance. Recall@5 is used for most M-BEIR test sets, except for Fashion200K and FashionIQ, where Recall@10 is applied.
Meanwhile, we divide the ten retrieval tasks into three sub-tasks — Single, Mixed, and Multi-Modal, based on the modalities of queries and candidates.

\noindent \textbf{Implementation Details.}
We primarily evaluate our method on the Qwen2-VL 7B model. During training, we prune the first $k$ layers of the MLLM and only fine-tune the remained shallow model. The first stage is conducted on 4 A800 GPUs with a batch size of 72 per GPU, using a learning rate of 1e-4 and LoRA parameters $r=128$, $\alpha=256$. In the second stage, the batch size is increased to 150 per 4 GPUs, and the learning rate is set to 3e-4, with the same LoRA settings. We observe that pruning $k=12$ layers yields saturated performance on the UMR task, offering a good trade-off between efficiency and accuracy. For self-distillation, we set ${\alpha_1=0.9, \alpha_2=0.1}$. For MAC-loss, we set the decay sparsity $\lambda=0.5$.

\subsection{Experiment Results}
\textbf{Main Results.} As shown in Table \ref{main table}, We report the performance of our method on the M-BEIR benchmark, where our model achieves consistently strong results across various retrieval tasks when the number of truncated layers is set to k = 12. Specifically, our approach surpasses the CLIP-based model \cite{wei2024uniir} by 3.6 points and outperforms the LamRA-Ret-2B \cite{liu2025lamra} baseline by 2.3 points. Meanwhile, MLLM-based models outperform CLIP-based models on the more complex Mixed- and Multi-Modal subtasks by 7.4 and 10.9 points, indicating that stronger multimodal understanding leads to better performance on more challenging retrieval scenarios. Our method balances efficiency and performance, aiming to maximize efficiency while minimizing the impact on retrieval accuracy.

\noindent \textbf{Efficiency Results.}~In Table \ref{efficient}, we present a comparative analysis with the existing 7B MLLM-based retrieval models. Since MMEmbed \cite{lin2025mmembed} adopts a different training strategy and backbone architecture, our main comparison is with the Qwen2VL-7B model. Our training was conducted under more limited GPU resources, both baselines use A100 (80G) GPUs, while we use A800 (80G), and we achieve a 4x reduction in memory usage. Additionally, our model reduces FLOPs by 57.3\% compared to fully layers fine-tuned baselines. 
Furthermore, we evaluate the inference speed across three sub-tasks. Using a single GPU with a batch size of 64, we measure the average number of samples processed per second. Our model improves inference speed by 95.8\%, effectively doubling throughput, while maintaining an average performance gap of only 1.8 points across all datasets. These improvements also make MLLM-based models more cost-effective for real-world applications.

\begin{figure}[htbp]
\vspace{-8pt}
  \centering
  \includegraphics[width=1.0\linewidth]{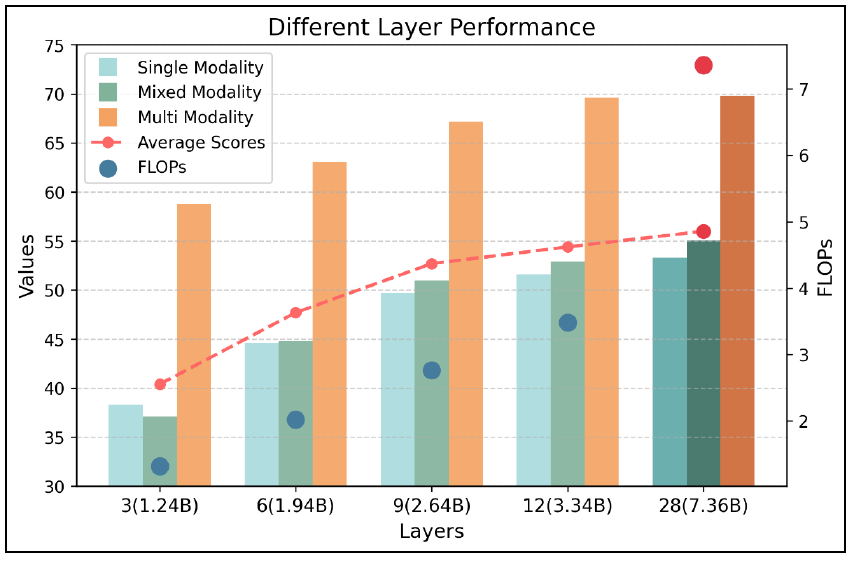}
  \caption{Performance visualization across layers. We display the average scores at selected layers and full models (2B and 7B) across three sub-tasks. Lines represent performance, and circles indicate FLOPs. The x-axis reflects the parameter scale at different layers.}
  \label{different layers}
  \vspace{-15pt}
\end{figure}

\noindent \textbf{Different Layers Performance.}~As shown in Figure \ref{different layers}, We present the performance, parameter count, and FLOPs for different values of k. 
\textbf{1).}~We offer a more flexible configuration space, where settings of k=9 to 12 generally offer a good balance between efficiency and performance. Compared to the 7B model in the final column, our approach reduces FLOPs by more than half while maintaining comparable performance.
\textbf{2).}~Regarding more tiny variants, a comparison across the first three columns reveals that when the number of layers exceeds a certain threshold, performance on the UMR task is significantly enhanced.
Our approach offers a better and selectable balance of performance and efficiency for 7B size. Additionally, it is fully compatible with the distilled models like Qwen2-VL-2B, allowing lightweight configurations by retaining only the layers most critical for retrieval performance.

\subsection{Ablation Study}
\textbf{Ablation of All Components.}
As presented in Table \ref{main ablation}, we report the results of our ablation studies. Our proposed methods, which address both architecture and learning strategy, have proven to be effective. Specifically, the self-distillation mechanism provides additional supervisory signal during training after layer-pruning, leading to performance gains of over 0.6 points across all three sub-tasks. Meanwhile, the MAC loss effectively mitigates the degradation of representational capacity and yields greater improvements, particularly in multi-modal sub-tasks. When combined, these two techniques result in an average performance increase of 1.3 points.

\noindent \textbf{Effectiveness of Dynamic Ratio in Pretraining.}
As shown in Table \ref{dynamic alpha}, we explore the impact of different weighting strategies for the contrastive loss ($\alpha_1$) and the self-distillation loss ($\alpha_2$) during pretraining. The Fixed setting maintains a constant ratio of 0.9/0.1 throughout training. In the Reverse setting, the ratio linearly shifts from 0.5/0.5 to 0.1/0.9, while the Dynamic setting increases it linearly from 0.5/0.5 to 0.9/0.1. Results indicate that the Dynamic strategy leads to improved pretraining performance. This suggests that assigning more weight to self-distillation in the early training stage—thus focusing on reconstructing semantic representations—followed by gradually shifting the focus to contrastive learning can yield better results. These findings support our analysis that the self-distillation loss helps recover shallow semantic representations, while the deeper layers of a well-trained teacher model already possess strong semantic capabilities that remain effective even without additional contrastive training.

\begin{table}
\vspace{-5pt}
  \caption{Ablation Study of All Components. We evaluate the impact of self-distillation and MAC loss on the UMR task across three sub-tasks.}
  \begin{tabular}{c c | c c c}
    \toprule
    Self-Distillation & MAC Loss & Single & Mixed & Multi \\
    \midrule
    $\times$ & $\times$ & 42.5 & 43.6 & 61.4 \\
    $\times$ & $\checkmark$ & 43.3 & 44.0 & 62.3 \\
    % $\checkmark$ & $\times$ & 42.5 & 43.1 & 61.8 \\
    % $\checkmark$ & $\checkmark$ & \textbf{43.1} & \textbf{44.3} & 61.9 \\
    $\checkmark$ & $\times$ & 43.6 & 44.2 & 62.1 \\
    $\checkmark$ & $\checkmark$ & \textbf{44.2} & \textbf{44.7} & \textbf{62.9} \\
  \bottomrule
  \label{main ablation}
\end{tabular}
\vspace{-15pt}
\end{table}

\begin{table}
 \vspace{-5pt}
  \caption{We compare the performance of different $\alpha_1$ and $\alpha_2$ settings under both a fixed ratio and two dynamic strategies: Reverse and Dynamic, which correspond to linearly decreasing and increasing ratios between the two, respectively.}
  \setlength{\tabcolsep}{4pt}
\begin{tabular}{c | c @{\hspace{3pt}} c @{\hspace{3pt}} c @{\hspace{3pt}} c}
    \toprule
     \multirow{2}{*}{\textbf{Loss}} & \multicolumn{2}{c}{\textbf{Image Retrieve}} & \multicolumn{2}{c}{\textbf{Text Retrieve}} \\
    \cmidrule(lr){2-3} \cmidrule(lr){4-5}
    & Flickr30k@5 & Coco@5 & Flickr30k@5 & Coco@5 \\
    \midrule
    Fixed & 91.8 & 65.0 & 95.8 & 73.9 \\
    Reverse & 82.2 & 60.2 & 89.6 & 69.8 \\
    \rowcolor{blue!10}
    Dynamic & \textbf{92.7} & \textbf{65.3} & \textbf{96.2} & \textbf{74.7} \\
  \bottomrule
\end{tabular}
\label{dynamic alpha}
\vspace{-5pt}
\end{table}

\begin{table}
\vspace{-5pt}
  \caption{Compare the effectiveness of modality-adaptively learning in more resource-constrained scenarios. We compare our method without MAC and reverse MAC, where the temperature coefficients for intra-group and inter-group samples are exchanged.}
  \begin{tabular}{c | c c c}
    \toprule
    Method & Single & Mixed & Multi \\
    \midrule
    w/o MAC Loss & 41.7 & 42.7 & 60.5 \\
    w/ Reverse MAC Loss & 41.8 & 42.3 & 60.7 \\
    \rowcolor{blue!10}
    w/ MAC Loss & \textbf{42.9} & \textbf{44.0} & \textbf{61.3} \\
  \bottomrule
\end{tabular}
\label{efficiency}
\vspace{-10pt}
\end{table}

\noindent \textbf{Effectiveness of Modality-Adaptive Learning Loss.}
In Table \ref{efficiency}, we simulate a more resource-constrained setting to evaluate the effectiveness of our MAC Loss. The model is trained on four GPUs with a reduced batch size of 90 per GPU. We observe that incorporating our contrastive learning loss consistently improves performance across three sub-tasks. In the second row, we replace the decayed temperature coefficient with an inter-modal coefficient while keeping the intra-modal temperature as norm. We refer to this variant as Diverse-Loss. Its performance remains similar to the setting without temperature decay, indicating that treating intra-group samples as harder negatives can bring performance gains, also validating our strategy's effectiveness in improving training efficiency under limited resources.

\begin{figure*}[htbp]
\vspace{-5pt}
  \centering
  \includegraphics[width=0.98\linewidth]{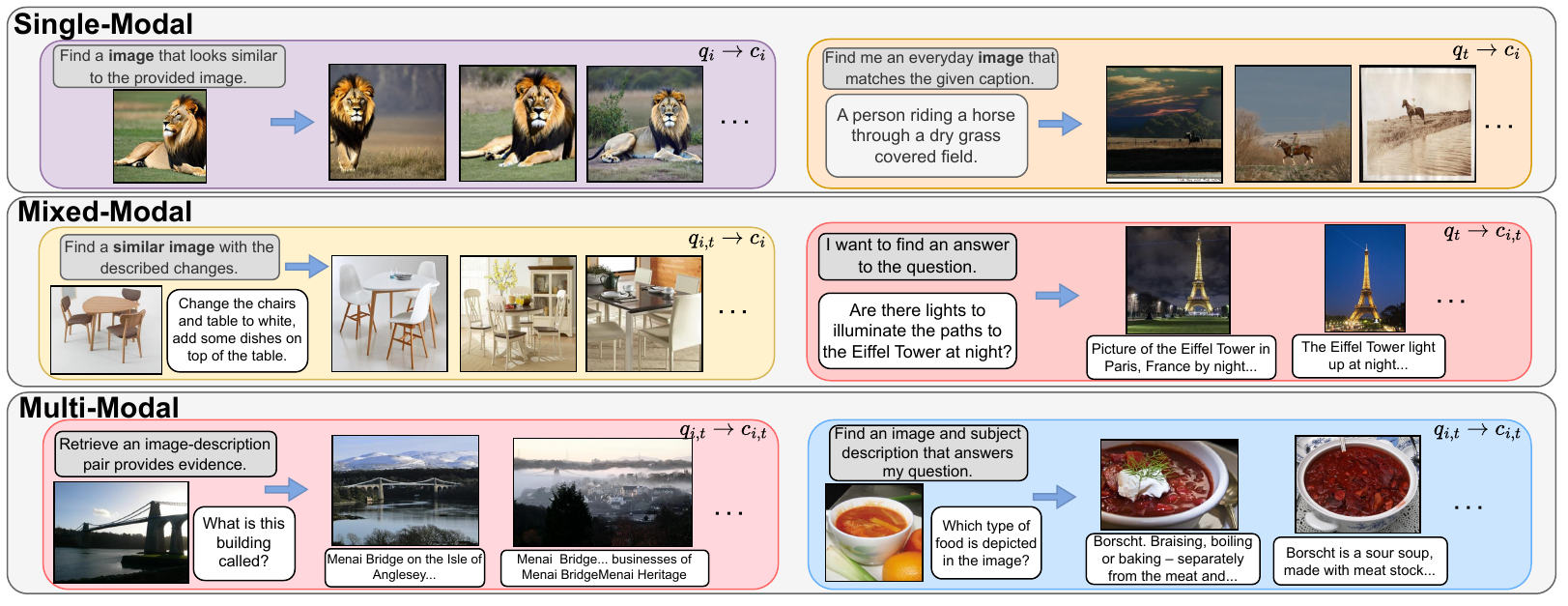}
  \caption{Some qualitative results. We present visualizations of representative cases from different retrieval tasks. The gray box indicates the input instruction used for retrieval, while the samples displayed to the right of the arrow are ranked by similarity. The first sample corresponds to the ground truth, and the remaining retrieved samples can also provide positive information.}
  \label{case-study}
  \vspace{-2pt}
\end{figure*}

% 消融1：不同layer与2b(FLOPS)

\subsection{Comparison with Token Compression Method}
We compare our method with token compression, another common approach for improving large model efficiency. Here, we use FastV \cite{chen2024image} as a baseline. Specifically, we apply the token compression technique to LamRA-Ret-7b and compare its performance with our method. It's worth noting that existing token compression methods mainly focus on image tokens and are generally ineffective for accelerating text inputs, we only compare the image token compression method in tasks with image modality input.

\vspace{-2pt}
\begin{table}
  \caption{Comparison with Token Compression Methods. We compare our method with the token compression method on three tasks involving image inputs.}
  \label{tab:freq}
  \begin{tabular}{c | c c c | c}
    \toprule
    \multirow{2}{*}{Method} & ${q_t \to c_i}$ & ${q_i \to c_t}$ & ${q_i \to c_i}$ & \multirow{2}{*}{FLOPs} \\
    \cmidrule(lr){2-4}
    & COCO & COCO & Nights & \\
    \midrule
    w FastV \cite{chen2024image} & 78.3 & 88.9 & 30.3 & 4.72 \\
    \rowcolor{blue!10}
    w PUMA & \textbf{80.3} & \textbf{91.1} & \textbf{31.5} & \textbf{3.48} \\
  \bottomrule
\end{tabular}
\vspace{-15pt}
\end{table}

Table \ref{efficiency} shows the results of our method and FastV \cite{chen2024image} on three types of retrieval tasks. Our approach consistently outperforms while requiring fewer than 0.62 FLOPs. In text-only retrieval, FastV provides no acceleration, highlighting the advantage of our layer-pruning strategy in enabling efficient MLLM-based UMR model.

% case-study
\subsection{Qualitative Results}
Figure \ref{case-study} presents retrieval cases based on different multimodal inputs and instructions. In addition to retrieving the ground truth (first column), the model retrieves many other relevant samples. Compared to fixed-modal inputs, these more complex settings better represent real-world retrieval needs. The model's ability to handle such cases demonstrates its wide applicability in practical scenarios.

\begin{figure}[htbp]
\vspace{-2pt}
  \centering
  \includegraphics[width=1.0\linewidth]{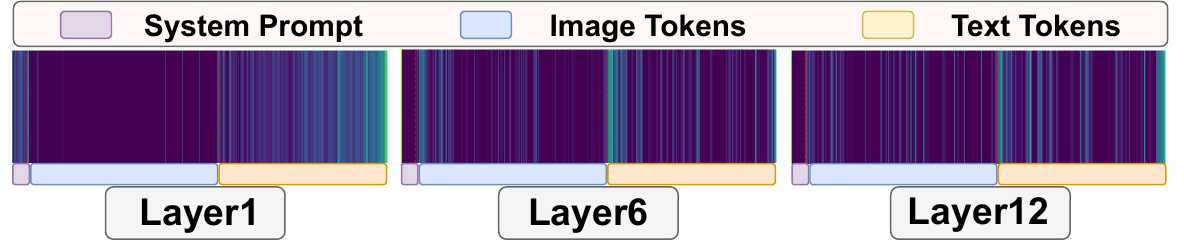}
  \caption{Visualization of the attention weights from the [RET] token to all other tokens after two-stage fine-tuning. The tokens are arranged from left to right as system prompt tokens, image tokens, and text tokens.}
  \label{attention2}
  \vspace{-10pt}
\end{figure}

Table \ref{attention2} shows the attention patterns of PUMA after training with zooming in [RET] token to other tokens. We can observe that the shifting focus of the [RET] token from initially attending solely to the text modality to gradually expanding its attention across a wider range of tokens, with a more concentrated focus on specific tokens as the layers progress, gradually aggregating information from all tokens to a few, and eventually to the RET token.

\section{Conclusion and Future Work}
In summary, we propose PUMA, an efficient unified multimodal retrieval framework improved from the model structural and learning perspective. From the structure perspective, we introduce a layer-pruned self-distillation method that keeps retrieving relevant shallow layers and uses the discarded deep layer as teacher model, creating a lightweight model with comparable performance. From the learning perspective, we tackle premature convergence in multimodal contrastive learning with a modality-adaptive learning loss, which adaptively samples hard negatives for each query based on its modality. PUMA reduces computation and memory costs while maintaining strong retrieval performance, making it well-suited for real-world UMR applications.

However, current MLLM-based UMR models do not show a clear advantage over CLIP-based models on single-modal tasks. Moreover, our method still presents performance limitation, future work can focus on these challenges. As an upstream task of Retrieval-Augmented Generation (RAG), UMR task enables the extension of RAG from a text-only paradigm to a multimodal setting, offering significant research value for the advancement of MM-RAG.

\newpage
\bibliographystyle{ACM-Reference-Format}
\bibliography{main}

\cleardoublepage

\appendix
\clearpage
{
\newpage
   \twocolumn[
    \centering
    \Huge
    \textbf{PUMA: Layer-Pruned Language Model for Efficient Unified
Multimodal Retrieval with Modality-Adaptive Learning}\\
    \vspace{0.5em}Supplementary Material \\
    \vspace{1.0em}
   ]
}
\section{More Details about M-BEIR Datasets}
\subsection{M-BEIR Dataset Composition}
To help understand the UMR task, in this section, we provide additional details of the M-BEIR dataset. The table presents a comprehensive overview of the constituent datasets included in each task. The M-BEIR benchmark encompasses eight types of retrieval tasks across ten datasets, comprising a total of 5.6 million candidate instances. This table is excerpted from the UniIR \cite{wei2024uniir}; for additional details, please refer to the original publication.

\begin{table}[h]
\centering
\caption{Summary of the M-BEIR benchmarks. M-BEIR has 8 tasks, 10 datasets with different modality input.
}
\setlength{\tabcolsep}{3mm}{
  \begin{tabular}{lllr}
    \toprule
    \textbf{Task} & \textbf{Dataset} & \textbf{Domain} & \textbf{\# Pool}\\
    \midrule
    \multirow{3}{*}{$q^t \to c^i$} & VisualNews & News & 542K \\
     & MSCOCO & Misc. & 5K \\
     & Fashion200K & Fashion & 201K \\
     \midrule
     \multirow{1}{*}{$q^t \to c^t$} & WebQA & Wiki & 544K \\
     \midrule
     \multirow{3}{*}{$q^i \to c^t$} & VisualNews & News & 537K \\
     & MSCOCO & Misc. & 25K \\
     & Fashion200K & Fashion & 61K \\
     \midrule
     \multirow{1}{*}{$q^i \to c^i$} & NIGHTS & Misc. & 40K \\
     \midrule
     \multirow{2}{*}{$q^t \to (c^i, c^t)$} & EDIS & News & 1M \\
     & WebQA & Wiki & 403K \\
     \midrule
     \multirow{2}{*}{$(q^i, q^t) \to c^t$} & OVEN & Wiki & 676K \\
     & InfoSeek & Wiki & 611K \\
     \midrule
     \multirow{2}{*}{$(q^i, q^t) \to c^i$} & FashionIQ & Fashion & 74K \\
     & CIRR & Misc. & 21K \\
     \midrule
     \multirow{2}{*}{$(q^i, q^t) \to (c^i, c^t)$} & OVEN & Wiki & 335K \\
     & InfoSeek & Wiki & 481K \\
     \midrule
    8 tasks & 10 datasets & 4 domains & 5.6M \\
    \bottomrule
  \end{tabular}
}

\label{mbeir}
\end{table}

\begin{table*}[ht]
\centering
\caption{Summary of the M-BEIR instructions. M-BEIR prepared four instructions for each dataset. We randomly select one instruction for display.
}
\resizebox{.86\textwidth}{!}{  % 控制表格的整体宽度
\setlength{\tabcolsep}{0.6mm}{
  \begin{tabular}{lll}
    \toprule
    \textbf{Task} & \textbf{Dataset} & \textbf{Instruction} \\
    \midrule
    \multirow{3}{*}{$q^t \to c^i$} & VisualNews & Based on the caption, provide the most fitting image for the news story. \\
    
    & MSCOCO & Show me an image that best captures the following common scene description.\\
     
     & Fashion200K & Based on the following fashion description, retrieve the best matching image.\\
     \midrule
     $q^t \to c^t $ & WebQA & Retrieve passages from Wikipedia that provide answers to the following question.\\
     \midrule
     \multirow{2}{*}{$q^t \to (c^i, c^t) $} & EDIS & Identify the news photo for the given caption.\\
      & WebQA & Find a Wikipedia image that answers this question.\\
     \midrule
     \multirow{3}{*}{$q^i \to c^t $} & VisualNews & Based on the shown image, retrieve an appropriate news caption. \\
     & MSCOCO & Find an image caption describing the following everyday image. \\
     & Fashion200K & Based on the displayed image, retrieve the corresponding fashion description.  \\
     \midrule
     $q^i \to c^i $ & NIGHTS & Which everyday image is the most similar to the reference image? \\
     \midrule 
     \multirow{2}{*}{$(q^i, q^t) \to c^t $} & OVEN & Retrieve a Wikipedia paragraph that provides an answer to the given query about the image.\\
     & InfoSeek & You have to find a Wikipedia segment that answers the question about the displayed image. \\
     \midrule 
     \multirow{2}{*}{$(q^i, q^t) \to c^i $} & FashionIQ  & With the reference image and modification instructions, find the described fashion look. \\
     & CIRR & I'm looking for a similar everyday image with the described changes. \\
    \midrule
    \multirow{2}{*}{$(q^i, q^t) \to (c^i, c^t) $} & OVEN & Determine the Wikipedia image-snippet pair that clarifies the entity in this picture. \\
    & InfoSeek & Determine the Wikipedia image-snippet pair that matches my question about this image. \\
    \bottomrule
  \end{tabular}
}
}

% \vspace{-2em}
\label{instruction}
\end{table*}

\subsection{M-BEIR Dataset Instructions}
As shown in Table \ref{instruction}, we present a subset of instructions used for unified multimodal retrieval, selecting one representative instruction from each dataset for illustration. The M-BEIR benchmark provides four diverse instructions per dataset, designed to cover different phrasings or intentions for the same retrieval task. In this section, we randomly select one instruction from the four for display in the table. During both training and evaluation, one instruction is randomly sampled from the available four for each sample and used as the input prompt, which encourages the model to generalize across various instruction formulations.

\section{More Experiment Results}
\subsection{Evaluate Our Method on LLaVA}
We also evaluate our method on the LLaVA-v1.5 \cite{liu2023visual}. As shown in Table~\ref{llava}, our method leads to a 57.8\% drop in FLOPs, while retaining most of the model’s capability when preserving 12 layers (k = 12). We follow most of the settings from MMEmbed \cite{lin2025mmembed}, training LLaVA-v1.5 on 8 A800 GPUs, while our method is trained on 4 A800 GPUs. Although MMEmbed is built upon LLaVA-Next \cite{liu2024llavanext}, it discards the cropping strategy, which leads to more than 1K image tokens. This significantly limits the batch size, weakening the effectiveness of contrastive learning and potentially causing out-of-memory errors during training. Therefore, we conduct our experiments directly on LLaVA-v1.5 to evaluate our method.

\begin{table}
  \caption{Evaluation results of our method on LLaVA.}
  \begin{tabular}{c | c c c | c}
    \toprule
    Method & Single & Mixed & Multi & Flops \\
    \midrule
    LLaVA & 50.0 & 51.3 & 66.9 & 10.91 \\
    \rowcolor{gray!20}
    w PUMA & 48.1 & 49.2 & 66.7 & \textbf{4.61} \textcolor[rgb]{0.0, 0.5, 0.0}{\small$\downarrow$ \textbf{57.8\%}} \\
  \bottomrule
\end{tabular}
\label{llava}
\end{table}

\subsection{Effectiveness of Different Distill Loss.}
In Table \ref{distill}, we compare different loss functions used for self-distillation when pretraining. We experiment with cosine similarity and KL divergence, also commonly used losses in knowledge distillation. Cosine similarity is applied to embedding tokens, while KL divergence is used to distill similarity scores from contrastive learning. For evaluation, we use Flickr30k \cite{plummer2015flickr30k} and COCO \cite{lin2014microsoft} after pre-training, following the same setup as LamRA. Results show that applying any distillation loss improves performance. Specifically, MSE outperforms the other two methods by an average of 1.1 points and brings a 4-point gain compared to training without a distillation loss. KL divergence performs better on text retrieval, while cosine similarity excels in image retrieval.

\begin{table}
 \vspace{-5pt}
  \caption{Comparison of Distillation Losses. We present the scores obtained after pre-training with different feature distillation loss functions. We evaluate the model on both image and text retrieval tasks.}
  \setlength{\tabcolsep}{4pt}
\begin{tabular}{c | c @{\hspace{3pt}} c @{\hspace{3pt}} c @{\hspace{3pt}} c}
    \toprule
     \multirow{2}{*}{\textbf{Loss}} & \multicolumn{2}{c}{\textbf{Image Retrieve}} & \multicolumn{2}{c}{\textbf{Text Retrieve}} \\
    \cmidrule(lr){2-3} \cmidrule(lr){4-5}
    & Flickr30k@5 & Coco@5 & Flickr30k@5 & Coco@5 \\
    \midrule
    w/o Distill & 88.3 & 62.1 & 90.3 & 69.5 \\
    \midrule
    Cosine Loss & 91.4 & 64.8 & 93.8 & 71.4 \\
    KL Loss & 91.2 & 64.2 & 94.6 & 72.0 \\
    \rowcolor{gray!20}
    MSE Loss & \textbf{91.8} & \textbf{65.0} & \textbf{95.8} & \textbf{73.9} \\
  \bottomrule
\end{tabular}
\label{distill}
\vspace{-5pt}
\end{table}

\subsection{Further Discussion about Layer-pruning}
Here, we further discuss the difference between layer-pruning and choosing a smaller model. First, our method is orthogonal to model size; for example, the same layer-pruning strategy can also be applied to a 2B-scale model such as Qwen2VL-2B. In addition, when applying our method to the 7B model, the resulting model at approximately k=6 layers has a parameter count comparable to that of Qwen2VL-2B. Under the same training setup, our approach achieves performance that is largely on par with the 2B model while maintaining lower FLOPs.

\subsection{Further Discussion about Modality-Separation and Modality-Gap}
We begin by presenting a theoretical analysis of modality separation as shown in Figure \ref{tSNE}. Similar findings have been observed in CLIP \cite{liang2022mind}, where embeddings—despite being generated by identical encoder architectures—tend to cluster in different "conical" regions. This suggests an inherent geometric separation between modalities. Another potential explanation lies in statistical differences: images and text vary significantly in structure, dimensionality, and density. Text is typically sparse and abstract, whereas images are dense and information-rich. As a result, their representations naturally diverge to reflect these underlying disparities \cite{li2021align}.

On the other hand, we further discuss the relationship between modality-separation and modality-gap. Some studies suggest that reducing the modality gap can lead to better retrieval performance \cite{role2025fill, fahim2024s}. Their focus is on cross-modal retrieval mismatches. For example, when the input is in the text modality but the retrieved results tend to lean toward the text rather than the image. In such cases, narrowing the modality gap can help alleviate this issue.
In contrast, our proposed MAC is designed to mitigate premature convergence of multimodal contrastive learning. Within a reasonable range of the hyperparameter $\lambda$, it does not aggravate the aforementioned retrieval mismatch problem.
We believe both directions represent promising research avenues for the UMR task. Moreover, a deeper exploration of the relationship between separation and gap would be also valuable for future work.

\begin{figure*}[htbp]
  \centering
  \includegraphics[width=0.9\linewidth]{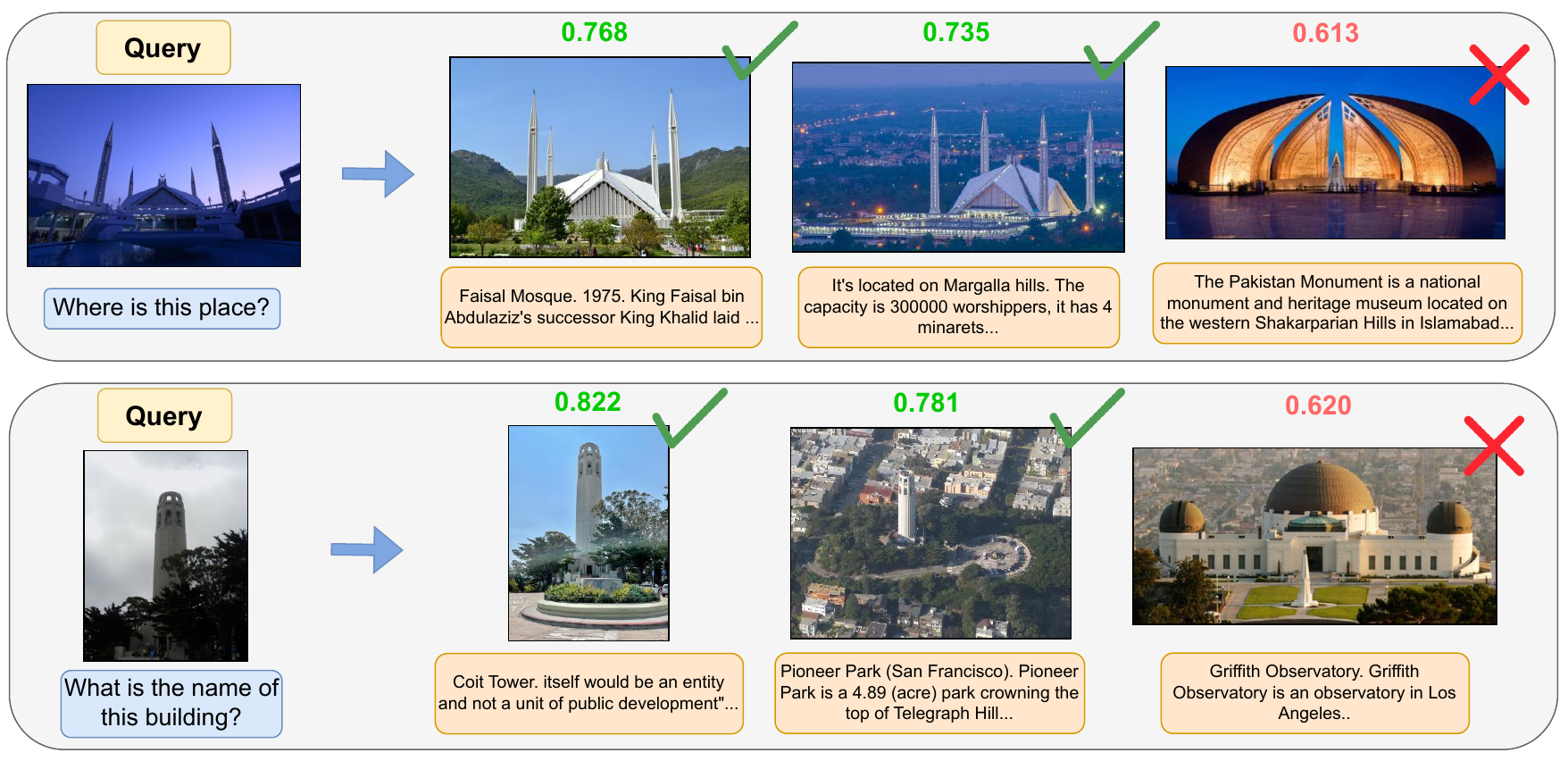}
  \caption{Visualization of Quantitative Results. We present visualizations of some representative examples to qualitatively assess retrieval performance. For each query, we display the top-3 retrieved candidates ranked by similarity score, with the ground truth consistently shown in the first column. The similarity score for each retrieved image is annotated above the corresponding candidate.}
  \label{case-study-appindx}
\end{figure*}

\subsection{More Qualitative Results and Analysis}
In Figure \ref{case-study-appindx}, we further visualized some retrieval examples and observed an interesting phenomenon during evaluation on the Oven dataset \cite{hu2023open}. Specifically, when the query pertains to human geography, there are cases where the similarity scores exhibit a sudden drop, where the top-ranked candidates have significantly higher similarity scores than the rest. Upon inspection, we found that these high-scoring candidates often provide valid and informative supplements to the query, even though they are not labeled as positive candidates in the dataset. The abrupt decline in similarity suggests that the remaining candidates are evidently irrelevant to the query. This observation may offer a potential criterion for distinguishing negative candidates, which could be particularly beneficial for downstream tasks such as retrieval-augmented generation (RAG). This bifurcation pattern suggests that human geography queries may inherently possess distinguishing characteristics – such as well-defined geopolitical boundaries, unique cultural identifiers, or specific geospatial relationships – that enable more discriminative relevance matching. From a modeling perspective, the observed sharp relevance decay implies that the learned representation space effectively captures domain-specific ontological structures, creating measurable separation between conceptually adjacent and disparate entities.

\end{document}